\documentclass[10pt,twocolumn,floatfix,aps,physrev,superscriptaddress]{revtex4-2}
\usepackage{amsmath,amssymb,amsthm,mathrsfs,amsfonts,dsfont,amstext} 
\usepackage{textcomp,pbox}
\usepackage[export]{adjustbox}
\usepackage{bm}
\usepackage{dcolumn,booktabs,url}
\usepackage[scaled]{helvet}
\usepackage{sansmath,gensymb}
\usepackage{tikz,graphicx,transparent,color}
\usepackage{multirow}
\usepackage[separate-uncertainty = true]{siunitx}
\usepackage{comment}
\usepackage{physics}
\usepackage{pdfpages}
\usepackage{array}
\usepackage{makecell}
\usepackage{tabularx}
\usepackage{dsfont}
\usepackage{braket}

\usepackage{tikz}
\usepackage{tikz-feynman}
\usetikzlibrary{arrows.meta,calc}
\tikzfeynmanset{every vertex/.style={dot, minimum size=1pt, inner sep=0pt}}

\newcolumntype{L}{>{\raggedright\arraybackslash}X}

\makeatletter
\AtBeginDocument{\let\LS@rot\@undefined}
\makeatother

\newcolumntype{C}[1]{>{\centering\let\newline\\\arraybackslash\hspace{0pt}}m{#1}}

\usepackage[colorlinks=true]{hyperref}
\usepackage{graphicx}
\usepackage{float}

\hypersetup{
     colorlinks   = true,
     citecolor    = red,
     linkcolor    = blue,
     urlcolor     = red     
}

\graphicspath{{./figs/}}

\newcommand{\fref}[1]{\text{Fig.}~\ref{#1}}

\usepackage{lipsum}
\usepackage{graphicx}
\usepackage{import}
\usepackage{xifthen}
\usepackage{pdfpages}
\usepackage{transparent}
\usepackage{svg}
\usepackage{adjustbox}
\usepackage{calc}
\graphicspath{{./figs/}}
\usepackage{physics}

\begin{document}

\title{Resonant Energy Transfer and Collectively Driven Emitters in Waveguide QED}

\newcommand{\AffCPH}{Center for Hybrid Quantum Networks (Hy-Q), Niels~Bohr~Institute, University~of~Copenhagen, Blegdamsvej 17, DK-2100 Copenhagen, Denmark}
\newcommand{\AffBochum}{Lehrstuhl f\"ur Angewandte Festk\"orperphysik, Ruhr-Universit\"at Bochum, Universit\"atsstra\ss e 150, 44801 Bochum, Germany}

\author{Cornelis Jacobus van Diepen$^\ddag$}
\thanks{These authors contributed equally to this work.}
\affiliation{\AffCPH{}}
\author{Vasiliki Angelopoulou}
\thanks{These authors contributed equally to this work.}
\affiliation{\AffCPH{}}
\author{Oliver August Dall'Alba Sandberg}
\affiliation{\AffCPH{}}
\author{Alexey Tiranov}
\altaffiliation{Present address: Chimie ParisTech, Université PSL, CNRS, Institut de Recherche de Chimie Paris, 75005 Paris, France}
\affiliation{\AffCPH{}}
\author{Ying Wang}
\affiliation{\AffCPH{}}
\author{Sven Scholz}
\author{Arne Ludwig}
\affiliation{\AffBochum{}}
\author{Anders S{\o}ndberg S{\o}rensen}
\author{Peter Lodahl}
\thanks{Email to: cornelis.diepen@nbi.ku.dk; lodahl@nbi.ku.dk}
\affiliation{\AffCPH{}}

\begin{abstract}
Waveguide quantum electrodynamics (QED) has opened a new frontier in quantum optics, which enables the radiative coupling of distantly located emitters via the spatially extended waveguide mode. This coupling leads to modified emission dynamics and previous work has reported the observation of increased intensity correlations (an antidip) when probing the resonance response of multiple emitters. However, the interference between independent emitters has been shown to lead to a similar response. Here, we directly observe resonant energy transfer between two distant quantum emitters by recording an antidip in the intensity correlations, $g^{(2)}(\tau)$, while driving only one of the emitters. Under the condition that only a single emitter is driven, the antidip in photon coincidences is a distinctive signature of emitter-emitter coupling, which enables the transfer of energy from the driven to the undriven emitter. Interestingly, the observed mechanism  is a long-range and waveguide-engineered version of resonant Förster transfer, which is responsible for the transport of energy between chlorophylls in the photosynthesis. Building on the established coupling, we demonstrate collective driving of the coupled emitter pair. Specifically, we control the relative driving phase and amplitude of the emitters and apply this collective excitation scheme to selectively populate the long-lived subradiant state. This results in suppressed emission, i.e. the peculiar situation where driving two emitters as opposed to one effectively reduces the probability of photon emission. Our work presents novel emission regimes and excitation schemes for a multi-emitter waveguide QED system. These can be exploited to deterministically generate emitter-emitter entanglement and advanced photonic states providing robustness against losses for photonic quantum computation and quantum communication. 
\end{abstract}

\maketitle 

\section{Introduction}
Radiative emission is not an immutable property of the emitter itself but can be modified by the environment, notably as described by the Purcell effect~\cite{Purcell1946}. This further entails that multiple coupled emitters have modified dynamics, as famously proposed by Dicke, leading to either enhanced (superradiant) or suppressed (subradiant) emission~\cite{Dicke1954, Gross1982}. In waveguide quantum electrodynamics (QED), emitters can couple to one another via a single propagating mode~\cite{Sheremet2023}. In nanophotonic waveguides with efficient radiative coupling (high $\beta$-factor)~\cite{Lodahl2015}, the otherwise weak emitter-emitter coupling can be strongly enhanced. In essence, emitters that are spatially separated by many optical wavelengths can be coupled, ultimately only limited by the propagation length of the waveguide. Consequently, the emitters can be addressed separately, i.e. excitation schemes can be implemented where the emitters can be driven either individually or collectively, where the latter refers to the coherent driving of all the emitters simultaneously. 

Collective effects in waveguide QED have been pioneered in the microwave domain with superconducting qubits, where the collective driving of waveguide-coupled emitters has been used to prepare dark states~\cite{Zanner2022}, 
to perform gates between giant atoms~\cite{Kannan2020}, and to generate two-dimensional photonic cluster states~\cite{OSullivan2024}. In the optical domain, waveguide-mediated coupling has been observed through modifications in lifetime measurements, which were supplemented with a proof-of-principle demonstration of emission dynamics that arise when collectively driving the emitters~\cite{Tiranov2023}. Control over the collective driving can enable selective excitation into a long-lived subradiant state leading to the generation of entanglement by dissipation~\cite{Gullans2012}. Subradiant states have also been proposed for enhanced efficiency of quantum transduction based on an increased interaction time~\cite{Elfving2019}. Furthermore, control over coupled emitters will enable the deterministic generation of multi-photon entangled states~\cite{Economou2010, Gonzalez-Tudela2015, Rubies-Bigorda2024}, enabling loss  mitigation in a quantum internet~\cite{Kimble2008, Wehner2018} or photonic quantum computation~\cite{Raussendorf2001}.

\begin{figure*}
 	\includegraphics[width=\textwidth]{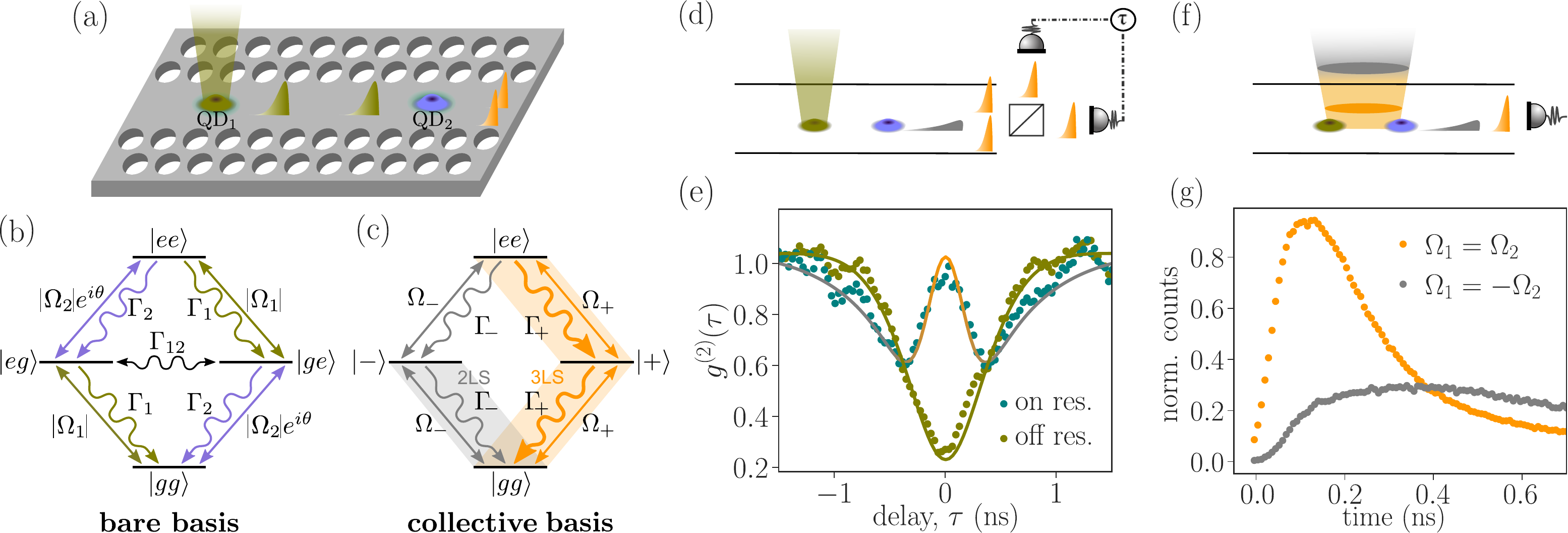}
	\caption{\textbf{Device schematic, energy diagrams and collective emission phenomena.} 
    (a) Schematic illustration of a photonic crystal waveguide with two quantum dot (QD) emitters, of which QD$_1$ is driven, and single-photon wavepackets transferred  between the QDs. 
    (b), (c) Energy level diagrams for two dissipatively-coupled quantum emitters in the (b) bare basis and (c) collective basis. The driving of the QDs is indicated  by two-sided straight arrows and decays by colored wavy arrows. In the bare basis, $|\Omega_i|$ is the absolute Rabi frequency for QD$_i$, and $\theta$ is the relative driving phase. The waveguide-mediated coupling between the emitters is indicated by the two-sided wavy arrow labeled $\Gamma_{12}$. The collective basis is in terms of the bright state, $\ket{+}$, and the dark state, $\ket{-}$, with their respective Rabi frequencies, $\Omega_\pm$, and decay times, $\Gamma_\pm$. In this basis, the system can conceptually be separated in a slowly decaying two-level system (highlighted in gray and labeled ``2LS") and a rapidly decaying three-level system (highlighted in orange and labeled ``3LS"). 
    (d) Schematic of two emitters in a waveguide with one continuously driven and a Hanbury Brown and Twiss setup which equally splits the scattered light into two detectors.
    (e) Two-photon intensity correlations, $g^{(2)}(\tau)$, for resonant (teal) and off-resonant (olive) QDs, obtained with the setup shown in (d). The laser frequency is in resonance with QD$_1$, and the power is such that the Rabi frequency is $\Omega_1 / 2 \pi = 0.25$ GHz $= 0.34$ $\Gamma_1 / 2 \pi$, while QD$_2$ is not driven (see Appendix~\ref{app:appa}). The solid lines show simulations based on independently extracted experimental parameters. The oscillations at longer times are an artifact caused by crosstalk between the channels of the time tagger. 
    (f) Schematic of a collectively driven pair of emitters in a waveguide and a single-photon detector. The pulsed laser field drives the emitters either in-phase (gray) or out-of-phase (orange).
    (g) Time-resolved emission intensity for in-phase (orange), $\Omega_1=\Omega_2$, and out-of-phase (gray), $\Omega_1=-\Omega_2$, pulsed excitation of the pair of resonant emitters, $\Delta_{12}/2\pi\approx 0$ GHz.}
	\label{fig:1}
\end{figure*}

In quantum optics, the second-order intensity correlation function, $g^{(2)}(\tau)$, with $\tau$ the time delay between photon detections, can be used to extract statistics and underlying properties of the emitters. An ideal single-photon source, for example, leads to anti-bunching~\cite{Kimble1977} by the absence of photon coincidences resulting in a dip in the correlation function ($g^{(2)}(0)=0$). This technique can be extended to multiple emitters, for which different driving schemes become applicable. By driving both in a pair of emitters, photon coincidences were measured in the form of an 'antidip', i.e. a peak superimposed on the coincidence dip (where $g^{(2)}(0) > 1/2$), for silicon-vacancy color centers~\cite{Sipahigil2016} and self-assembled quantum dots~\cite{Kim2018, Grim2019, Hallett2024}. These observations merely prove that the emitters are correlated, but cannot be taken as a distinctive signature of emitter-emitter coupling. In fact, an antidip has been reported without the modification of the radiative lifetime, which reveals the absence of radiative coupling between the emitters~\cite{Koong2022}. Despite the emitters being independent, the antidip arises due to correlations induced by path erasure via measurement of the first photon~\cite{Cygorek2023}. Recently, bunched emission was reported from incoherently excited closely-spaced quantum dots, coupled to a low Q cavity~\cite{Kim2024}.

In this article, we experimentally and theoretically study waveguide-mediated dipole-dipole coupling between two quantum emitters, and realize two remarkably different emission regimes depending on the excitation scheme. 
In the first regime, we drive continuously only one of the emitters yet measure photon coincidences, which may be captured by the phrase \textit{``drive one, get two"}. This is a consequence of the radiative coupling, which ensures that a single driven two-level emitter can result in the emission of two simultaneous photons by transfer of energy from the driven to the undriven emitter. We theoretically model the intensity correlations and find that, given that only a single emitter is driven, the observed antidip in $g^{(2)}(\tau)$ is a distinctive signature of coupling between the emitters. In the second regime, the emitters are collectively driven by pulsed excitation and the relative phase of the drive is controlled. In particular, when the emitters are driven out of phase, they are predominantly excited into the subradiant state. This subradiant state is characterized by suppressed decay, and thus results in emission that can essentially be expressed as \textit{``drive two, get none"}. Note that both regimes have in common that there is little population in the state $\ket{ee}$ (both emitters excited) and the suppressed decay of the subradiant state plays a fundamental role. 

\section{The platform and collective phenomena}
The experimental platform, schematically visualized in \fref{fig:1}(a), consists of a photonic crystal waveguide (PCW) with a pair of InGaAs quantum dot (QD) emitters inside. Each emitter consists of a ground, $\ket{g}$, and excited, $\ket{e}$, state and form the coupled four-level system as depicted in \fref{fig:1}(b). For each of the emitters, the probability is high that a photon is emitted into the guided mode, i.e. $\beta_i=\gamma^{wg}_i/\Gamma_i$ is close to unity~\cite{Arcari2014, Javadi2018}, where $\gamma^{wg}_i$ is the decay rate into the waveguide and $\Gamma_i$ the total QD$_i$ decay rate, $i=1,2$. The spacing between the emitters corresponds to a coupling phase, $\phi_{12} \approx N \pi$ with $N$ an integer, which results in a predominantly dissipative coupling, $\Gamma_{12}=\sqrt{\gamma^{wg}_1 \gamma^{wg}_2}$~\cite{Tiranov2023}. The emitters are driven from free space, and photons emitted into the guided mode are coupled out of the chip via collection ports at both ends of the waveguide. The emission frequencies can be tuned both electrically and magnetically, where the latter is used to control the detuning between the emitters. 

To directly observe energy transfer between the individual emitters, only one QD is continuously driven, and the scattered light is split into two detectors, as schematically shown in \fref{fig:1}(d). Figure~\ref{fig:1}(e) shows the obtained $g^{(2)}(\tau)$ with the emitters resonant (teal points) and far off-resonant (olive points). Far from resonance, with the detuning $\Delta_{12}/2\pi$ larger than the indvidual emitter linewidths, $\Gamma_1/2\pi = 0.73$ GHz and $\Gamma_2/2\pi = 0.79$ GHz, we observe clear antibunching, $g^{(2)}(0) = 0.24$, consistent with scattering from a single emitter. On the contrary, by tuning the two emitters into resonance, $\Delta_{12}/2\pi \approx 0$, a clear antidip with $g^{(2)}(0) = 0.94$ is observed. The observed photon coincidences while driving a single emitter are a direct consequence of the resonant transfer of energy from the driven to the undriven emitter by coupling via a shared optical mode.

Going beyond the steady-state regime, dynamics with enhanced and suppressed emission can occur for the coupled QDs. To directly access these, we implement pulsed collective driving of the coupled emitters, as shown by the schematic in \fref{fig:1}(f). For this driving scheme, the relative driving strength and phase play a key role, as they can be tailored to selectively drive into e.g. the (anti)symmetric collective state as indicated with ($\Omega_-$) $\Omega_+$ in \fref{fig:1}(c). Lifetime measurements of the collectively excited emitters are shown in \fref{fig:1}(g) where the two emitters are resonant with one another. When the emitters are driven in phase, $\Omega_1 = \Omega_2$, a bright emission burst is observed. In contrast, for out-of-phase driving, $\Omega_1=-\Omega_2$, the emission is initially strongly suppressed, and only at later time reduced emission occurs. 
This directly demonstrates how to control the collective emission from coupled quantum emitters. 

\section{Energy transfer and photon coincidences}
\label{sec:transf}
\begin{figure}
    \includegraphics[width=\columnwidth]{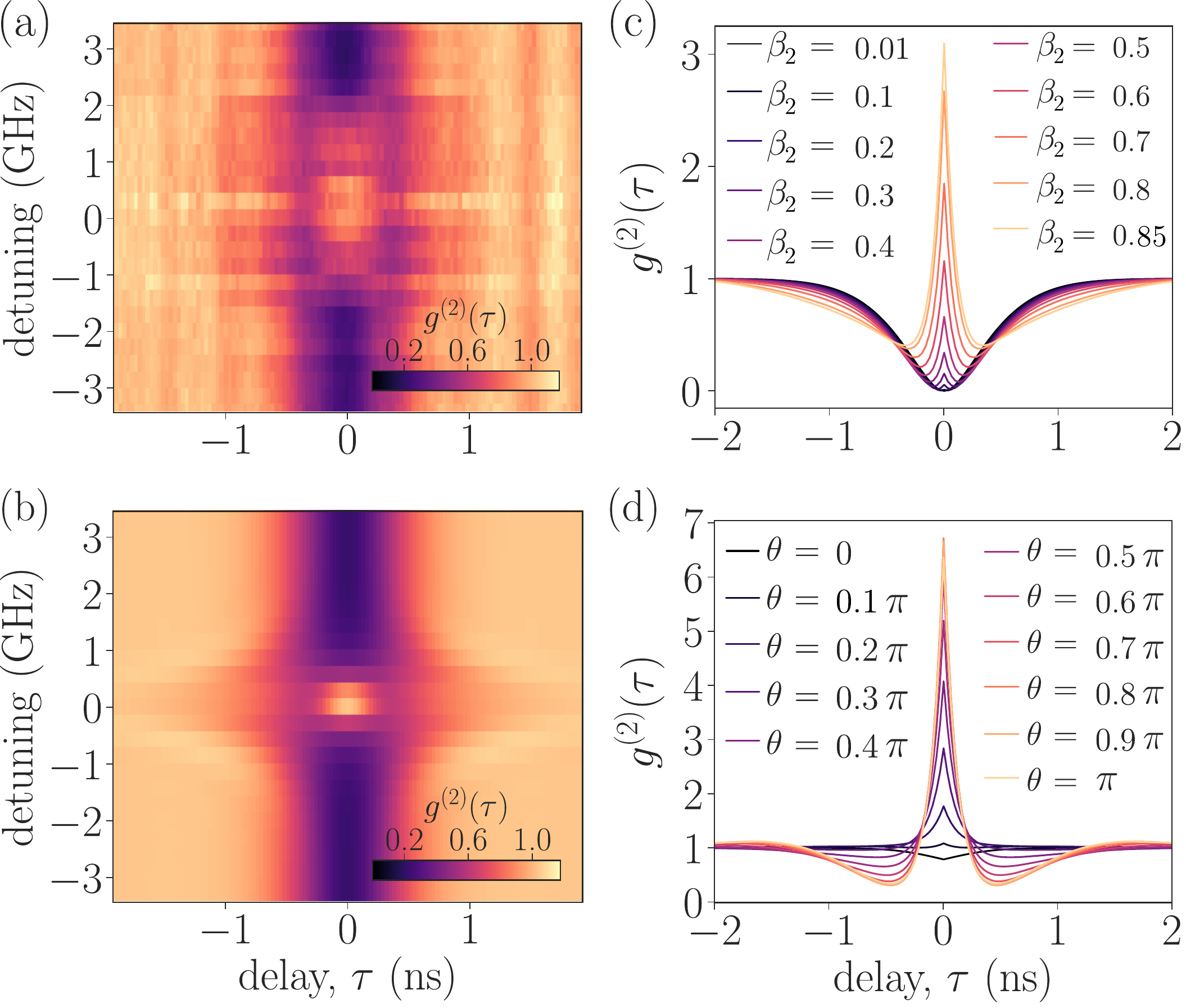}
    \caption{\textbf{Photon coincidences from coupled emitters.}
    (a) Measured second-order correlation function, $g^{(2)}(\tau)$, as a function of detuning, $\Delta_{12}/2\pi$, between the two quantum emitters while resonantly driving one emitter. 
	(b) Simulated $g^{(2)}(\tau)$ as a function of detuning corresponding to (a) using the experimental parameters and including detection jitter. 
    (c), (d) Simulations for $g^{(2)}(\tau)$ using experimental parameters and varying (c) the coupling strength of the second emitter to the waveguide mode, $\beta_2$, while driving only QD$_1$, and (d) varying the relative driving phase, $\theta$, while equally strongly driving both emitters, $|\Omega_1|=|\Omega_2|$.
    }
	\label{fig:2}
\end{figure}
An intuitive understanding of resonant energy transfer and the relation to photon coincidences can be obtained from the energy level diagram in the bare basis, shown in \fref{fig:1}(b). Let us consider the transfer as a sequential process. First, the driving of QD$_1$ brings population from the collective ground state, $\ket{gg}$, into the state $\ket{eg}$, as shown by the double arrow labelled $\Omega_1$. Next, via the radiative coupling indicated by $\Gamma_{12}$, energy in the form of a photon is transferred from the state $\ket{eg}$ into the state $\ket{ge}$. Finally, by further driving QD$_1$, the system is driven from $\ket{ge}$ into the doubly-excited state $\ket{ee}$, from which emission results in photon coincidences. 

Alternatively, the coupled emitters can be described in the collective basis for the single-excitation subspace, $\ket{\pm}=\frac{1}{\sqrt{2}}\left(\ket{eg} \pm \ket{ge}\right)$, shown in the energy level diagram in \fref{fig:1}(c). For purely dissipative coupling, $\ket{+}$ and $\ket{-}$ decay with superradiant and subradiant rates, $\Gamma_+$ and $\Gamma_-$, respectively. In addition, $\ket{ee}$ decays with rate $\Gamma_+$ ($\Gamma_-$) into $\ket{+}$ ($\ket{-}$), from which the subsequent decay results in photon coincidences. These photons predominantly come from the three-level system highlighted with the orange box in \fref{fig:1}(c) and result in an antidip in $g^{(2)}(\tau)$.

Another feature observed in the measured $g^{(2)}(\tau)$, shown in \fref{fig:1}(e), is a dip, reminiscent of a single two-level emitter, albeit temporally broadened. This broadened dip hints at the emission being dominated by an effective two-level system formed by the ground and subradiant state, highlighted with the gray box in \fref{fig:1}(c). For this system, the slow equilibration set by the subradiant decay rate would result in temporal broadening. Intuition for the formation of such an effective two-level system can be obtained from the quantum Zeno effect~\cite{QuantumZeno1977}. In this context, decay is interpreted as a measurement by the environment and the system is continuously projected back to the ground state, essentially preventing population from going into the superradiant and doubly-excited state. 

Explicitly, the rates of photon emission from the superradiant, $\ket{+}$, and subradiant state, $\ket{-}$, are given by $R_\pm = \Gamma_\pm p_\pm$, where $p_+$ ($p_-$) is the superradiant (subradiant) population. If the subradiance is not perfect, $\Gamma_- > 0$, and given a weak resonant drive, the difference in decay rates between $\Gamma_+$ and $\Gamma_-$ results in a difference in the populations $p_{\pm} = \frac{\Omega^2}{\Gamma_{\pm}^2}$, such that the rate of photon emission through the super- and subradiant states is $R_\pm = \frac{\Omega^2}{\Gamma_\pm}$. Assuming that the coupling to other modes is weak, this shows that, despite the suppressed decay, there are more photons coming from the subradiant state. Thus, the ground and subradiant state indeed effectively form a slowly decaying two-level system. It further shows that population accumulates in the slowly decaying $\ket{-}$ state. Interestingly, this mechanism can be used to deterministically generate entanglement, which is enabled by the dissipation of the emitters~\cite{Gonzalez-Tudela2011}. It is noteworthy that this scheme entangles two distant entities through local operations on only one of them.

An analytical expression for the photon coincidences can be obtained from a perturbative model. In terms of Feynman diagrams, $g^{(2)}(\tau)$ for a pair of super- and subradiant emitters, with decay rate $\Gamma_1=\Gamma_2=\Gamma$, and with one driven resonantly ($\Omega_1=\Omega$ and $\Omega_2=0$), to leading order in $\Omega$, is
\begin{equation}
g^{(2)}(\tau) = \left|
\resizebox{!}{0.45 cm}{
\begin{tikzpicture}[scale=1, baseline = (current bounding box.center),line width = 1 pt]
    \begin{feynman}
     \vertex (a) at (0,0);
        \vertex [right=1cm of a] (b);
        \vertex at ($(b) + (1cm, 2cm)$) (c); 
        \vertex [right=1cm of c] (d); 
        \vertex [below=2cm of d] (e);
        \vertex [right=1cm of e] (f);
        \vertex at ($(f) + (1cm, 2cm)$) (g); 
        \vertex [right=1cm of g] (h);
        \vertex [below=2cm of h] (i);
        \diagram* {
            (a) -- [photon] (b),
            (b) -- [double distance=1 pt, edge label=\huge{\(\Omega\)},purple] (c),
            (b) -- [fermion,draw opacity = 0, purple] (c),
            (c) -- [photon] (d),
            (d) -- [fermion,edge label' = \huge{$\Gamma_-$}] (e),
            (e) -- [photon] (f),
            (f) -- [double distance=1 pt, edge label=\huge{\(\Omega\)},purple] (g),
            (f) -- [fermion,draw opacity = 0,purple] (g),
            (g) -- [photon] (h),
            (h) -- [fermion, edge label' = \huge{$\Gamma_-$}] (i),
        };
    \end{feynman}
\end{tikzpicture}}
+ 
\resizebox{!}{0.9 cm}{
\begin{tikzpicture}[scale=1, baseline = (current bounding box.center), line width = 1pt]
    \begin{feynman}
     \vertex (a) at (0,0);
        \vertex [right=1cm of a] (b);
        \vertex at ($(b) + (1cm, 2cm)$) (c);
        \vertex [right=1cm of c] (d); 
        \vertex at ($(d) + (1cm, 2cm)$) (e);
        \vertex [right=1cm of e] (f);
        \vertex at ($(f) - (0cm, 2cm)$) (g);
        \vertex [right=1cm of g] (h);
        \vertex [below=2cm of h] (i);
        \diagram* {
            (a) -- [photon] (b),
            (b) -- [double distance=1 pt, edge label=\huge{\(\Omega\)},purple ] (c),
            (b) -- [fermion,draw opacity = 0,purple ] (c),
            (c) -- [photon] (d),
            (d) -- [double distance=1 pt,edge label=\huge{\(\Omega\)},purple ] (e),
            (d) -- [fermion,draw opacity = 0,purple] (e),
            (e) -- [photon] (f),
            (f) -- [fermion, edge label = \huge{$\Gamma_+$}] (g),
            (g) -- [photon] (h),
            (h) -- [fermion, edge label = \huge{$\Gamma_+$}] (i)
        };
    \end{feynman}
\end{tikzpicture}}
\right|^2
\label{eq: feynman}
\end{equation}
where $\Gamma_\pm = \Gamma \pm S$, with  $S= \sqrt{\Gamma_{12}^2 - \Delta_{12}^2}$, are the super- and subradiant decay rates. The horizontal axis represents time, whilst the vertical axis represents excitation number. The squiggly, horizontal lines represent the evolution of the system under the system's time propagator -- e.g. the effective Hamiltonian in the single-excitation subspace. For the diagrams, in accordance with the discussion above, it is assumed that the drive is predominantly into the subradiant state. The first diagram corresponds to an alternating sequence of excitation and decay, which results in a dip similar in character to a single-photon source. In contrast, the second diagram corresponds to two subsequent excitations followed by two decays, which results in an antidip. 

The equivalent normalized second-order correlation function is 
\begin{equation}
g^{(2)}(\tau) = \left| c+ c_- e^{-\Gamma_- \tau/2} + c_+ e^{-\Gamma_+ \tau/2}\right|^2,
\end{equation}
where $|c| = 1 + \mathcal{O}(\Omega^2)$ ensures the appropriate normalization. The constant term is a result of the steady state after the system has evolved from the first quantum jump. The $c_-$ term corresponds to the process of excitation from the ground state $\ket{gg}$, predominantly to the subradiant state $\ket{-}$, decay (with rate $\Gamma_-$) into $\ket{gg}$ and the subsequent re-excitation and decay. Both these terms are captured by the first Feynman diagram in eqn.~\eqref{eq: feynman}. They dominate at $\tau \gg 1/\Gamma_+ $ and correspond to the broad dip feature in \fref{fig:1}(e) and \fref{fig:2}. The $c_+$ term corresponds to a two-step excitation into the state $\ket{ee}$, and from there two-step decay with the first decay predominantly into the state $\ket{+}$ and the subsequent decay into $\ket{gg}$. This term is responsible for the narrow (width $\sim \Gamma_+^{-1}$) antidip feature. The antidip can in the ideal case and the limit of weak driving become arbitrarily high, i.e. the photons are strongly bunched. The theoretical model is further detailed in Ref.~\cite{Sandberg2024}.

The experimentally measured $g^{(2)}(\tau)$ for resonant emitters, shown in \fref{fig:1}(e), can, for delay times $|\tau| \ge 0.4$ ns, near the location of the minima in the ``W"-shape, be modeled with a broadened single-emitter dip with decay rate $\Gamma_{\text{dip}}/2\pi = 0.31(4)$~GHz, which matches with the subradiant lifetime measured in previous work~\cite{Tiranov2023}. The measured antidip can be modeled with a two-sided exponential decay with a decay rate $\Gamma_{\text{adip}}/2\pi = 0.52(2)$~GHz. This value is lower than the previously reported superradiant lifetime, due to detector jitter and spectral diffusion. Details and \fref{fig:SM_dip_antidip_independent_fits} with the fitted data are provided in the Appendix. 

In addition, we experimentally characterize and numerically simulate the antidip behavior as a function of detuning between the emitters. Figure~\ref{fig:2}(a) shows the experimentally acquired second-order intensity correlations. The data clearly show that the single-emitter antibunching dip transforms into an antidip as the emitters are brought into resonance. The numerically simulated data, shown in \fref{fig:2}(b), which is based on the full model that includes spectral diffusion and detection jitter, shows good agreement with the experimental data. We note that for a pair of coupled emitters driven by a spectrally narrow laser, modulation of the intensity correlations can occur. For example, detuning only the (not) driven emitter away from resonance with the laser can (decrease) increase the antidip height (see \fref{fig:SM_g2_diffusion} in the Appendix).

To further clarify the relation between radiative coupling and photon coincidences, we simulated $g^{(2)}(\tau)$ for two resonant emitters with various, $\beta_2$, i.e. various coupling strengths of the second emitter to the waveguide, while always driving only the first one. The results are shown in \fref{fig:2}(c), and demonstrate that the antidip is an effect of the high $\beta$-factor, thus of the radiative coupling between the emitters via the waveguide mode. Bunching, $g^{(2)}(\tau) > 1$, can be found in the simulations while this was not observed in the experiment, which is attributed to a combination of detection time jitter, finite driving power and various emitter imperfections such as pure dephasing, spectral diffusion and imbalanced decay rates. Altogether, the simulations corroborate that the observation of an antidip for a single driven emitter is a distinctive signature of the two emitters being coupled. Moreover, they confirm that transfer of energy from the driven to the undriven emitter only occurs for coupled emitters.
\begin{figure*}
	\includegraphics[width=\textwidth]{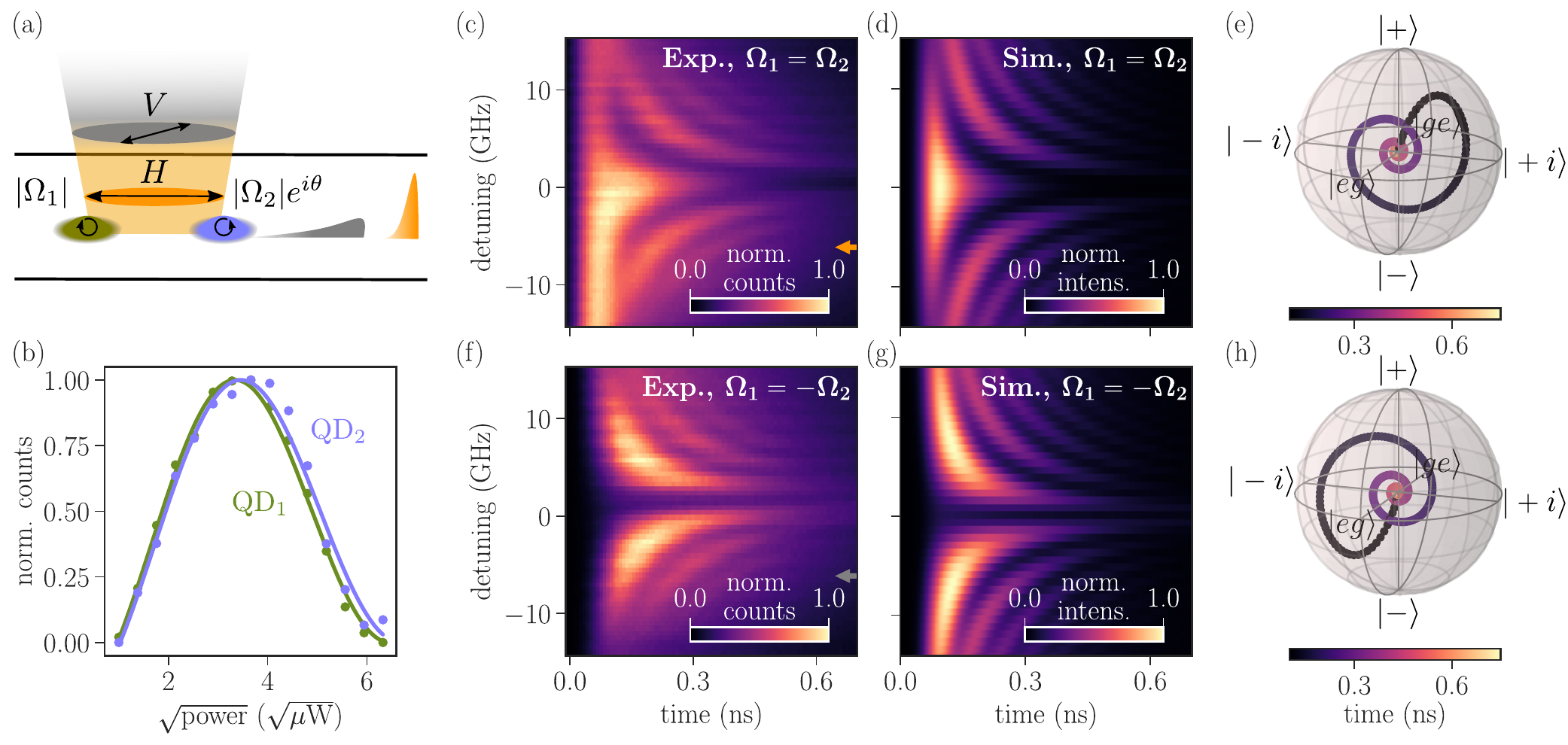}
	\caption{ \textbf{Control over super- and subradiant emission dynamics.} (a) Schematic illustration of pulsed collective excitation of two emitters. A relative driving phase is induced by the polarization of the driving field with respect to the dipole moments of the emitters. (b) Rabi oscillations for each emitter as a function of the square root of the power. Dots are data points and the solid lines are fits (see Appendix~\ref{app:appe}). (c), (f) Time-resolved emission as a function of detuning between the emitters for (c) in-phase driving, $\Omega_1=\Omega_2$, and (f) out-of-phase driving, $\Omega_1=-\Omega_2$. (d), (g) Numerical simulations corresponding to (c) and (f), respectively. (e), (h) Bloch spheres based on simulations of the state evolution in the single-excitation subspace for (e) in-phase and (h) out-of-phase driving. The detunings are indicated by the arrows in panels (c) and (f).
    }
	\label{fig:3}
\end{figure*}

In the case of collectively driving the emitters, the relative driving phase, $\theta$, is of importance. Figure~\ref{fig:2}(d) shows numerically simulated $g^{(2)}(\tau)$ as a function of that phase. For in-phase driving of the emitters, the antidip disappears, while for out-of-phase driving, the antidip height is increased as compared to driving only one emitter. This is intuitively explained by the fact that the driving phase controls the populations in the $\ket{+}$ and $\ket{-}$ states. Note that, these results are distinctively different from previous works reporting $g^2(0) \approx 1$~\cite{Sipahigil2016, Kim2018, Grim2019, Hallett2024, Koong2022} for a pair of emitters without relative driving phase coherence. These numerical results thus show the impact of phase coherence and its control when collectively driving emitters.

\section{Collective excitation}
Now we return to the implementation of collective driving. For multiple emitters, the driving Hamiltonian is
\begin{equation}
H_d = \frac{1}{2} \sum_m |\Omega_m| (e^{i \theta_m} \sigma_m^+ + e^{- i \theta_m} \sigma_m^-),
\end{equation}
where $\sigma_m^\pm$ are the Pauli raising and lowering operators for emitter $m$. The Rabi frequency, $|\Omega_m|$, and the driving phase, $\theta_m$, are given by
\begin{equation}
\label{eqn:Rabiterms}
|\Omega_m| e^{i\theta_m}= \frac{E_{0,m}}{\hbar} \langle e_m | \vec{\epsilon}_m \cdot \vec{d}_m | g_m \rangle ,
\end{equation}
with $E_{0,m}$ the electric field strength, $\vec{\epsilon}_m$ the polarization of the field, and $\vec{d}_m$ the transition dipole moment for emitter $m$. For a single emitter, the driving phase is typically ignored because it is a global phase. However, for the excitation of multiple emitters a relative driving phase is important. 

From eqn.~\eqref{eqn:Rabiterms} it follows, via the term $\langle e_m | \vec{\epsilon}_m \cdot \vec{d}_m | g_m \rangle$, that a specific polarization results in a certain driving phase. The polarization can thus be operated to selectively excite a collective state. In our experiment, the emitter transitions have orthogonal circular dipole moments as they have opposite magnetization~\cite{Warburton2013}. One emitter is thus optimally addressed by left-hand and the other by right-hand circularly polarized light, as illustrated in \fref{fig:3}(a). A horizontally (vertically) polarized driving field thus excites the emitters in-phase (out-of-phase). Alternative implementations for collective driving include using a shaped beam, e.g. a doughnut beam previously demonstrated with organic dye molecules~\cite{Trebbia2022}, or employing a spatial light modulator.

In order to equally excite both emitters, the Rabi frequencies should be made equal. Figure~\ref{fig:3}(b) shows Rabi oscillations for both emitters for a fixed polarization of the excitation laser. The nearly overlapping curves show that the Rabi frequencies are practically equal, i.e. $|\Omega_1| = |\Omega_2|$, for the respective polarization. For these measurements, the emission from both emitters was collected separately by detuning the emitters and spectrally filtering the emission. The beam position was optimized beforehand to equalize the driving field strengths for both emitters. Full maps for the driving as a function of polarization are shown in \fref{fig:SM_waveplatemaps} and \fref{fig:SM_Rabimap} in the Appendix. The former is taken at fixed power while the emission intensity is measured and the latter is based on a Rabi oscillation measurement for each polarization setting.

We employ control over the driving phase and strength to perform time-resolved emission measurements (lifetime measurements). A pulsed laser with power corresponding to a Rabi drive with a pulse area of $\pi/4$ for both emitters was used. At this power, the population in the $\ket{ee}$ state is low, thus the emission is dominated by that from the single-excitation subspace. Due to the coupling of the emitters, a strong dependence of the emission on the relative driving phase is expected. The energy level diagram in \fref{fig:1}(c) shows that either driving in phase, $\Omega_- = \frac{1}{\sqrt{2}} \left( \Omega_1 - \Omega_2 \right) = 0$, or out of phase, $\Omega_+=\frac{1}{\sqrt{2}} \left( \Omega_1 + \Omega_2 \right)=0$, results in selective excitation of either the superradiant or subradiant state, respectively.

Figure~\ref{fig:3}(c) and (f) show lifetime measurements as a function of detuning with the emitters driven (c) in phase, $\Omega_1=\Omega_2$, and (f) out of phase, $\Omega_1=-\Omega_2$. At resonance, as shown in \fref{fig:1}(g), the in-phase excitation results in a bright burst, due to the excitation of the superradiant state. The out-of-phase excitation leads to strongly suppressed emission, which demonstrates excitation of the subradiant state. With the emitters detuned, both lifetime measurements show multiple oscillations with a decaying amplitude. These oscillations are a consequence of the detuning, which results in a phase in the single-excitation subspace as $\ket{eg} + e^{i(\theta+\Delta_{12})t} \ket{ge}$. Essentially, the system evolves between super- and subradiant, thus the emission intensity oscillates with a declining amplitude due to the decreasing excitation.

The emission for opposite driving phase reaches its maximum at a later time than for the in-phase driving. This is explained by the fact that for the prior case, the subradiant state is populated first. After evolution into the superradiant state, the emission increases. For the latter case, with in-phase driving, the superradiant state is populated directly, thus emission is maximal earlier. This effect is visible in the Bloch spheres in \fref{fig:3}(e) and (h). These show the state evolution in the single-excitation subspace for in-phase and out-of-phase excitation, respectively. The state evolutions in (e) and (h) are based on simulations for the detunings indicated by the arrows in \fref{fig:3}(c) and (f), respectively. The numerically simulated time-resolved emission in \fref{fig:3}(d) and (g), which include spectral diffusion and jitter, show good agreement with the experimental results in (c) and (f), respectively. Together, these results demonstrate collective driving of the coupled emitters and their selective excitation into either predominantly the superradiant $\ket{+}$ or subradiant $\ket{-}$ state. \\

\section{Conclusion}
In summary, we directly observe the transfer of energy in the form of a photon between two individual emitters. The transfer manifests as an antidip in the photon coincidences while driving only one emitter. We argue that the observed antidip is a distinctive signature of emitter-emitter coupling using  an analytical expression and numerical simulations for the intensity correlations, $g^{(2)}(\tau)$. In addition, we perform pulsed collective excitation, and demonstrate control over the relative driving phase with polarization as the control knob. We utilize the collective excitation to access unexplored dynamics of the coupled emitters, and in particular to directly drive the system into the long-lived subradiant state. 

Selective excitation into the subradiant state enables a variety of applications. It allows direct access to decoherence-free subspaces~\cite{Zanner2022}, which are relevant for the reduction of information loss in open systems~\cite{Paulisch2016}. Furthermore, subradiant states can be used for photon storage as shown with dense atomic clouds~\cite{Ferioli2021}, or for efficient mirrors as implemented with ultracold atom arrays~\cite{Rui2020}.Interestingly, the demonstrated excitation schemes lead to deterministically generated entanglement via the dissipative coupling of the emitters~\cite{Gonzalez-Tudela2011, Reiter2012}. When combined with the spin degree of freedom~\cite{Warburton2013}, this naturally forms a path to spin-spin entanglement~\cite{Gullans2012} and quantum gates~\cite{Dzsotjan2010}. In turn, that enables the deterministic generation of advanced photonic cluster states~\cite{Economou2010}, which are key for robust encoding of photonic qubits, of relevance in measurement-based quantum computing~\cite{Raussendorf2001} and for the quantum internet~\cite{Wehner2018}. 

\section*{Acknowledgements}
The authors thank Bj\"{o}rn Schrinski for contributions to the theory at the beginning of this project, and Leonardo Midolo for contributions to device fabrication. Funding: The authors gratefully acknowledge financial support from Danmarks Grundforskningsfond (DNRF 139, Hy-Q Center for Hybrid Quantum Networks). O.A.D.S. acknowledges funding from the European Union’s Horizon 2020 research and innovation program under the Marie Skłodowska-Curie grant agreement no. 801199. S.S., and A.L. acknowledge financial support of the BMBF QR. N project 16KIS2200, QUANTERA BMBF EQSOTIC project 16KIS2061, as well as DFG excellence cluster ML4Q project EXC 2004/1. Competing interests: P.L. is founder of the company Sparrow Quantum, which commercializes single-photon sources. The other authors declare no competing interests.

\bibliography{references}

\clearpage
\onecolumngrid
\appendix

\section{Driving only one}
\label{app:appa}
\begin{figure}[h!]
 	\includegraphics[width=0.5\textwidth]{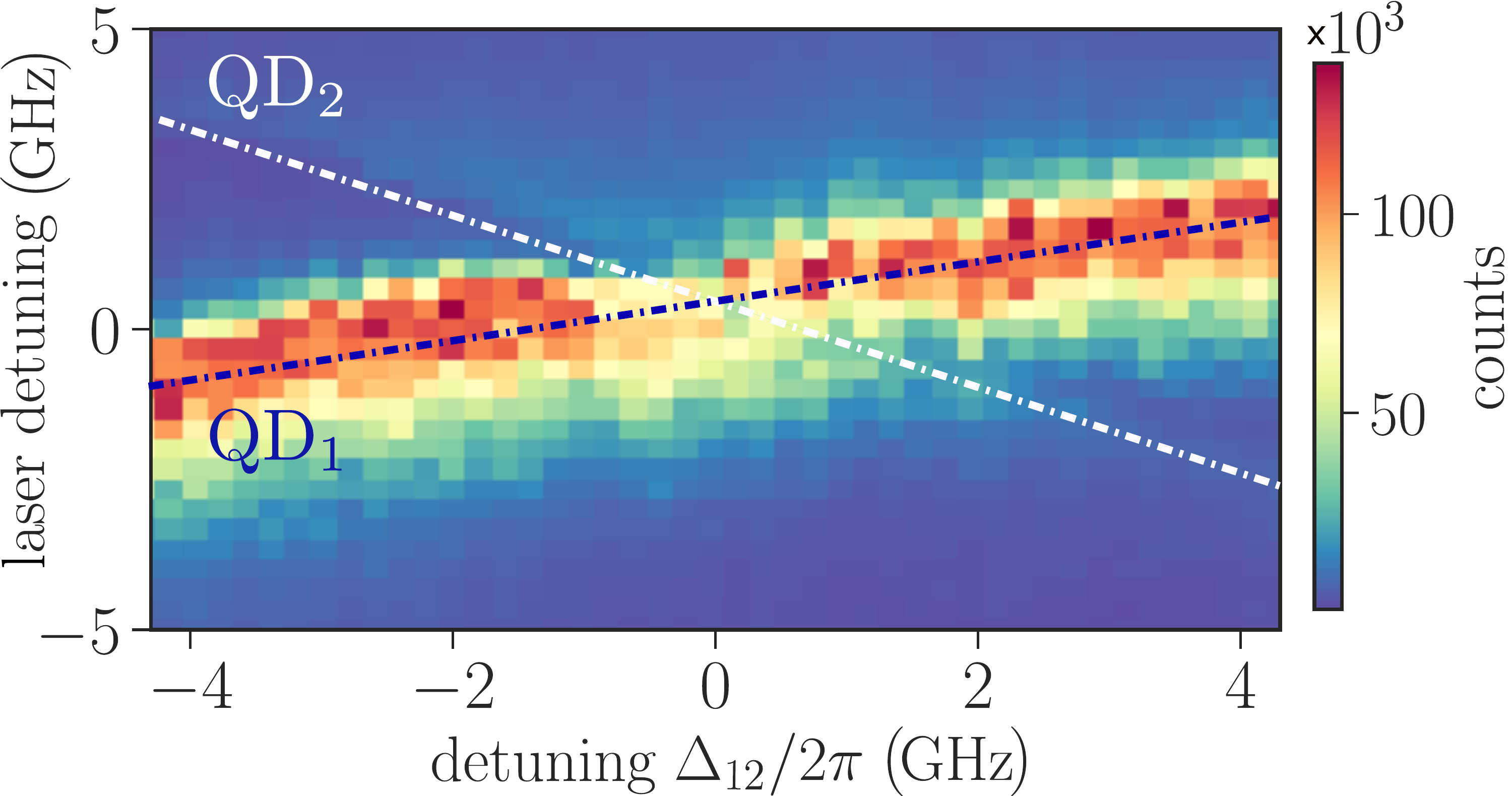}
	\caption{ 
   Resonance fluorescence spectrum while driving QD$_1$, as a function of the detuning between the emitters, $\Delta_{12}/2\pi$, achieved via a static magnetic field tuning the Zeeman splitting, and the laser detuning, controlled with a static electric field tuning the dc Stark shift. The blue (white) dash-dotted line is the expected resonance with the optical transition of QD$_1$ (QD$_2$) based on spectroscopy in resonant transmission, see Ref.~\cite{Tiranov2023}.
   }
	\label{fig:SM_RF}
\end{figure}

To assess whether only one emitter is driven, the resonance fluorescence spectrum is measured as a function of the detuning between the emitters, $\Delta_{12}/2\pi$, and the detuning with respect to the laser. The detuning of the laser is expressed as the frequency difference with respect to the frequency where the two emitters are resonant with another. The voltage was scanned as an efficient alternative for scanning the laser frequency. Figure~\ref{fig:SM_RF} shows the result, where the blue (white) dash-dotted line shows the resonance of the laser with QD$_1$ (QD$_2$). The data shows that QD$_1$ is driven while the driving of QD$_2$ is well suppressed, which is achieved by a combination of spatial and polarization selectivity. Another feature in the resonance fluorescence is the decrease of the intensity at the resonance between the two emitters ($\Delta_{12}/2\pi$ = 0). This is explained by a combination of the quantum Zeno effect, suppressing population in the superradiant state, and the suppressed decay of the subradiant state. 

\section{Correlations in low-frequency noise}
Spectral diffusion can impact the height of the antidip. To investigate this we perform numerical simulations for the case when driving only QD$_1$ with the parameters extracted experimentally. The effect of spectral diffusion on the correlations is assessed by scanning the detunings between the driving laser and the two QDs ($\Delta_1/2\pi$, $\Delta_2/2\pi$). 
\begin{figure}[h!]
 	\includegraphics[width=0.5\textwidth]{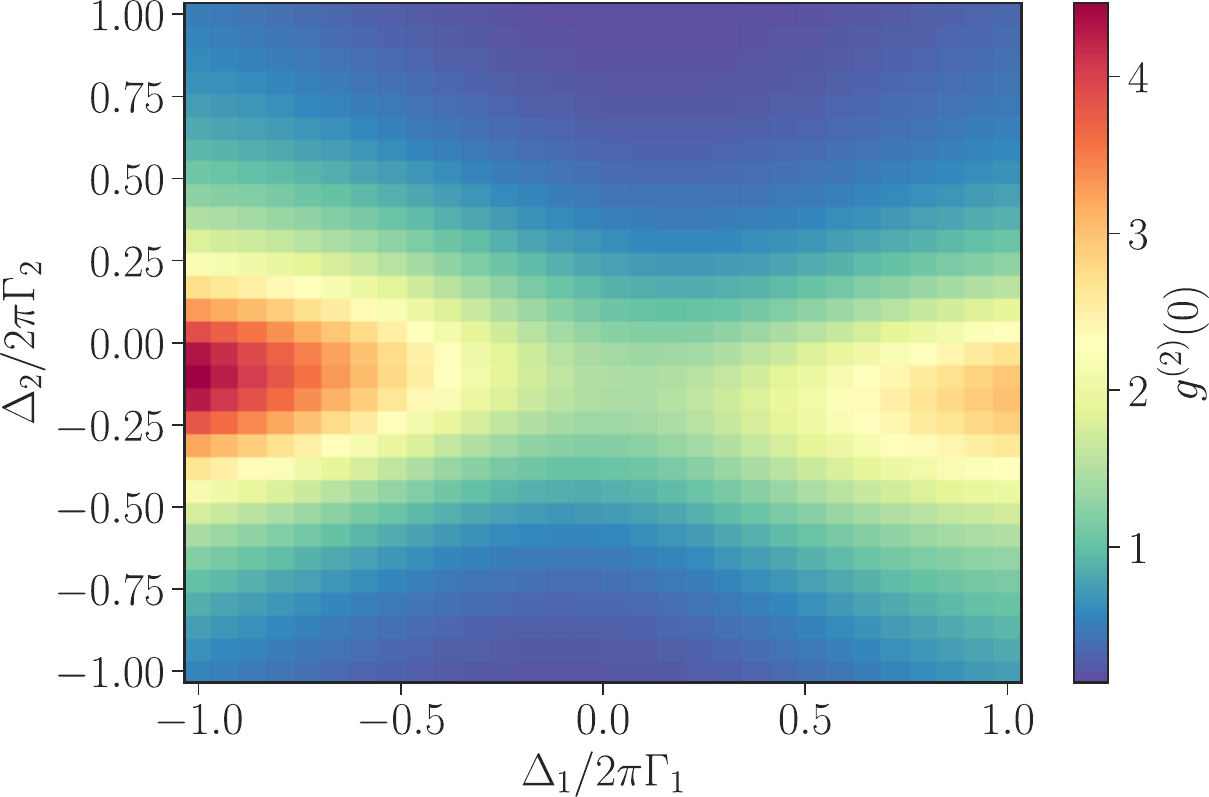}
	\caption{Dependence of $g^{(2)}(0)$ on the detuning between the laser and the driven QD$_1$ ($\Delta_1/2\pi$) and the non-driven QD$_2$ ($\Delta_2/2\pi$). For this simulation the experimental parameters are used.
   }
	\label{fig:SM_g2_diffusion}
\end{figure}

Figure~\ref{fig:SM_g2_diffusion} reveals that the antidip peak height varies strongly as it shows the highest values for large detunings between the laser and the driven emitter QD$_1$, whilst the non-driven QD$_2$ remains on resonance with the excitation frequency. This indicates that particular spectral diffusion on the driven emitter can result in an increased antidip.

\section{Extracting the super- and subradiant decay rates}
\begin{figure}[h!]
\centering
 	\includegraphics[width=.67\textwidth]{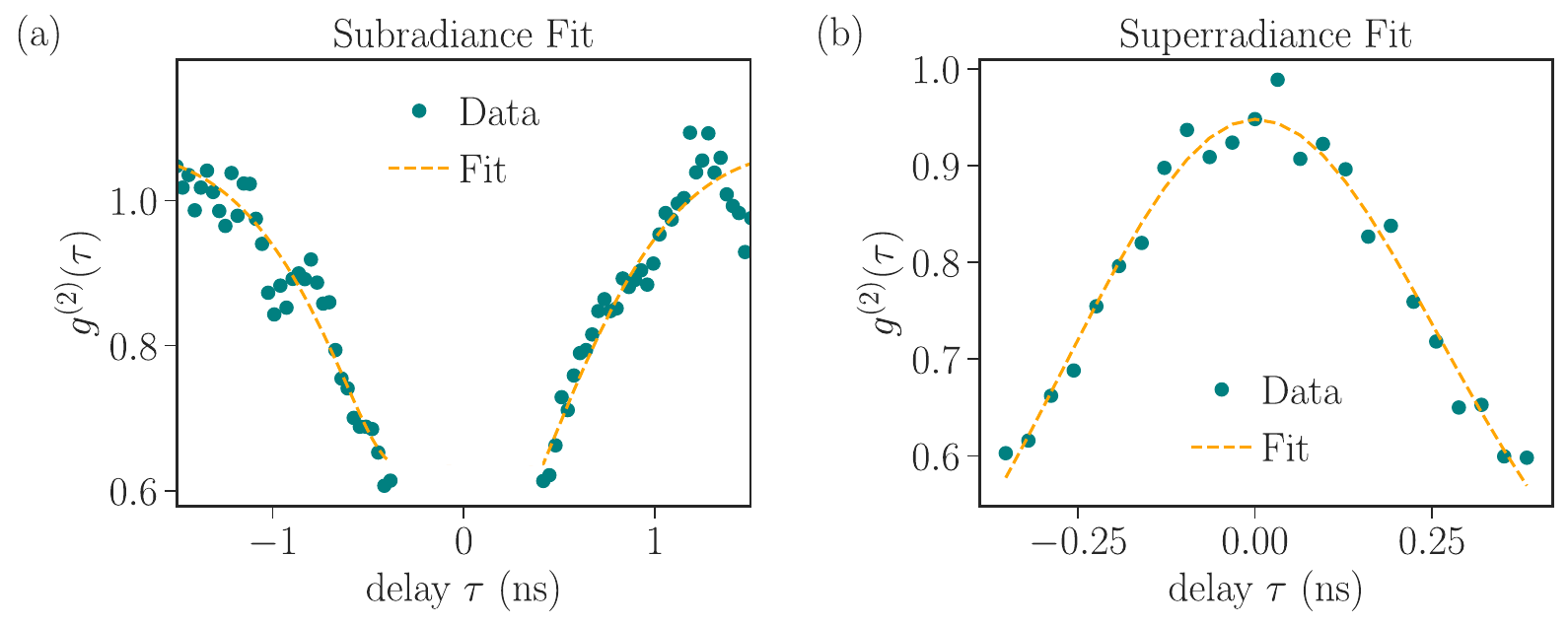}
	\caption{Independent fits of the two-photon intensity correlations, $g^{(2)}(\tau)$, for resonant QDs (see \fref{fig:1}(e), teal), with (a) a broadened single-emitter dip for the data related to the subradiant emission and (b) a two-sided exponential for the antidip related to the superradiant emission. }\label{fig:SM_dip_antidip_independent_fits}
\end{figure}

Considering the interpretation for the second-order intensity correlations described in Section~\ref{sec:transf}, the $g^{(2)}(\tau)$ for two QDs in resonance can be split in two time windows: the emission at longer delay times, related to subradiance, and the emission at short delay times, attributed to superradiance, as shown in \fref{fig:SM_dip_antidip_independent_fits}. More specifically, for delay times $\abs{\tau}\ge 0.4$ ns, the data can be effectively modeled using a broadened single-emitter dip described by 
\begin{equation}
g^{(2)}(\tau) = 1 - A e^{-\eta \tau} \left[ \cos{(\mu \tau)} + \frac{\eta}{\mu} \sin{(\mu\tau)}\right],
\end{equation}
where
\begin{equation}
\mu = 2 \pi \sqrt{\Omega^2 + \left(\frac{\Gamma_- - 2 \Gamma_d}{4}\right)^2}, \hspace{10pt} 
\eta = 2\pi \big(\frac{3\Gamma_- + 2 \Gamma_d}{4}\big).
\end{equation}
The antidip for delay times $-0.4\leq\tau\leq 0.4$ ns can be modeled with a two-sided single exponential. To account for the instrument response in the fits, each of the models is convolved with a Gaussian centered approximately at zero with a standard deviation determined by the time jitter of the two detectors as $\sigma = 0.35\sqrt{2}/\sqrt{8\log{2}}$, where $0.35$ ns is the measured full width at half maximum of the detector jitter. The fits yield $\Gamma_{dip}/2\pi = 0.31(4) $ GHz and $\Gamma_{adip}/2\pi = 0.52(2)$ GHz, as written in the main text.

\section{Waveplate maps and polarization control}
\begin{figure}[h!]
 	\includegraphics[width=\columnwidth]{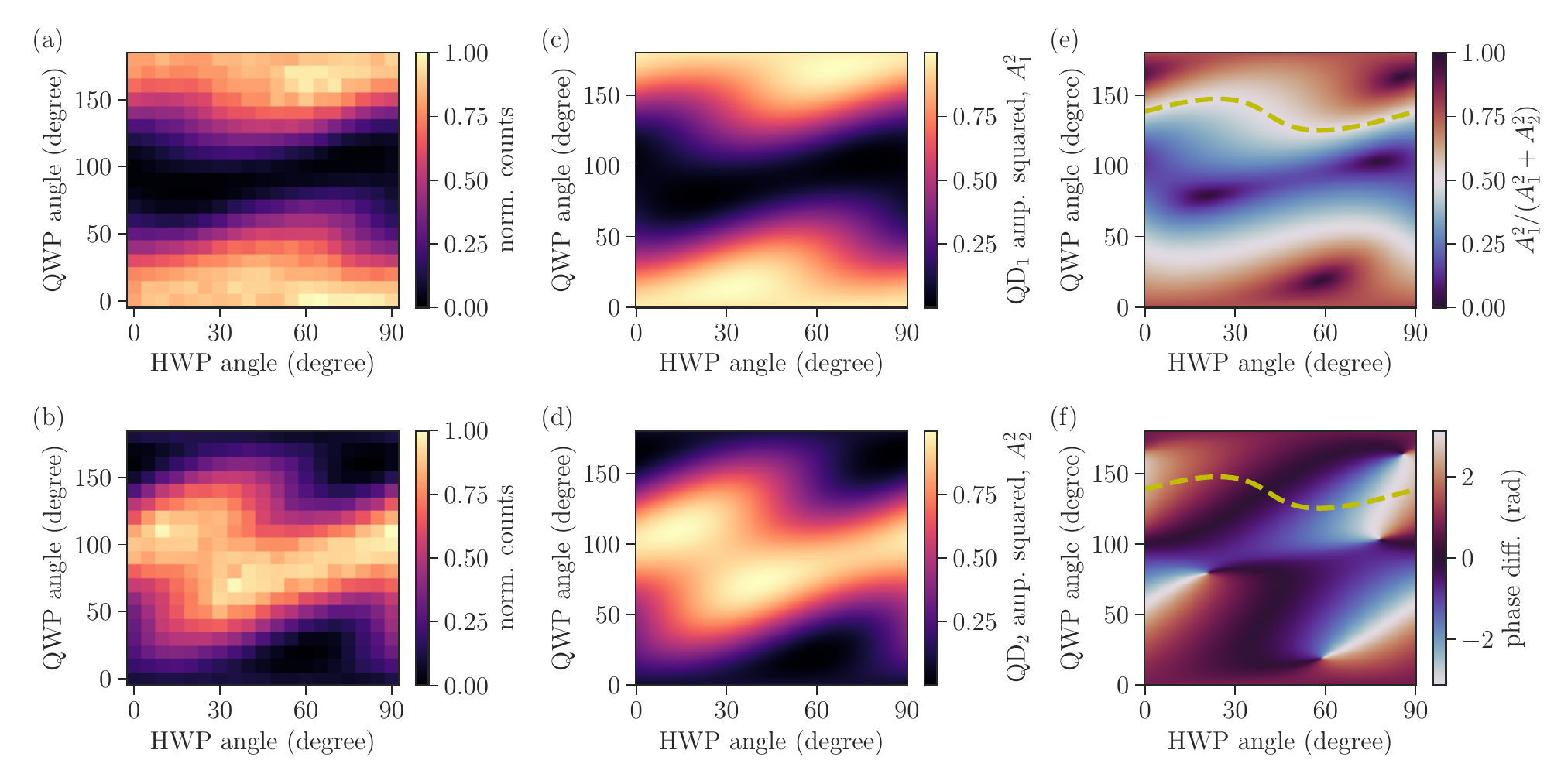}
	\caption{
    Intensity and phase as a function of waveplate rotation angles. (a), (b) Measured normalized counts for QD$_1$ and QD$_2$, respectively. (c), (d) Simulated amplitude squared $A_i^2\propto |\Omega_i|^2$ for QD$_1$ and QD$_2$, respectively. (e) Relative QD$_1$ amplitude squared, i.e. $A_1^2/(A_1^2 + A_2^2)$. (f) Rabi phase difference between QD$_1$ and QD$_2$. 
    The yellow dashed line traces out the contour $A_1 = A_2$ in (e), and is added in (f) to indicate the change in phase difference along that contour.
    }
	\label{fig:SM_waveplatemaps}
\end{figure}

The counts emitted from QD$_1$ and QD$_2$ for pulsed excitation and as a function of quarter-wave and half-wave plate rotation angles are shown in \fref{fig:SM_waveplatemaps}(a), and (b), respectively. The two intensity maps appear complementary to each other, i.e. where the counts are high for one, they are low for the other, and vice versa, which indicates that the dipoles are orthogonal. For this calibration, the detuning between the emitters is intentionally set to be relatively large, i.e. $>$\SI{5}{\giga\hertz}, such that emission from both can be collected separately by using an etalon filter (linewidth of $3$ GHz) and switching the emission frequencies electrically into and out of the filter. Experimentally, the polarization of the excitation laser, initially linearly polarized, is adjusted through computer-controlled rotation of a half-wave and a quarter-wave plate. Note that there is an arbitrary offset for both rotation angles due to the mounting of the waveplates. 

The emitted intensity for each of the emitters follows $I_i = \eta_{col} \gamma^{wg}_i p_{e,i}$ with $\eta_{col}$ the collection efficiency, $\gamma^{wg}_i$ the decay rate into the waveguide, and $p_{e,i}$ the excited state population for QD$_i$. If the drive power is in the linear regime of the Rabi oscillation, then $I_i \propto |\Omega_i|^2$, thus the normalized intensity in the waveplate maps corresponds to a relative Rabi amplitude squared. This is further confirmed by the agreement with the waveplate maps, shown in \fref{fig:SM_Rabimap}, which are based on direct measurements of the polarization dependence of Rabi oscillations as a function of driving power.

The intensity maps are for both emitters fitted with a model for the polarization based on the Jones formalism. This model calculates the absolute value squared of the inner product between the Jones vector for the transition dipole and that for the laser polarization, which we denote as $A_i^2$ and which is proportional to $|\Omega_i|^2$. The fitted maps are shown in \fref{fig:SM_waveplatemaps}(c), and (d), and show good agreement with the experimentally measured counts shown in \fref{fig:SM_waveplatemaps}(a) and (b), respectively. From the fitted maps, a relative amplitude squared for QD$_1$ is extracted, given by $A_1^2/(A_1^2 + A_2^2)$ shown in \fref{fig:SM_waveplatemaps}(e), and a phase difference, $\theta = \theta_1 - \theta_2$, shown in \fref{fig:SM_waveplatemaps}(f). The latter corresponds to the relative driving phase, and follows from the arguments of the inner products between the Jones vector for the laser and that for the respective transition dipoles. 

In addition, contour curves can be obtained, such as shown by the yellow dashed trace plotted in \fref{fig:SM_waveplatemaps}(e), along which the Rabi amplitude for both emitters remain equal to one another, i.e. $A_1=A_2$ which implies $|\Omega_1| = |\Omega_2|$, i.e. they are driven equally strongly. The remaining degree of freedom in the drive of both emitters is the relative phase. Thus, adjusting both wave plates to follow the yellow trace shown in \fref{fig:SM_waveplatemaps}(f) corresponds to only changing the relative driving phase. From parametrization of such a contour, an effective control knob for the phase is acquired. The change in phase along the contour is not constant, but from interpolation the required changes in the wave plates are obtained to achieve the desired change in phase. A similar approach can be used to control the relative drive power while fixing the phase difference. For simultaneous control over the relative drive phase and power, both knobs can be combined, which effectively constitutes a full map between the polarization and the drive parameters.

\section{Waveplate maps from Rabi oscillations}
\label{app:appe}
\begin{figure}[h!]
 	\includegraphics[width=.67\textwidth]{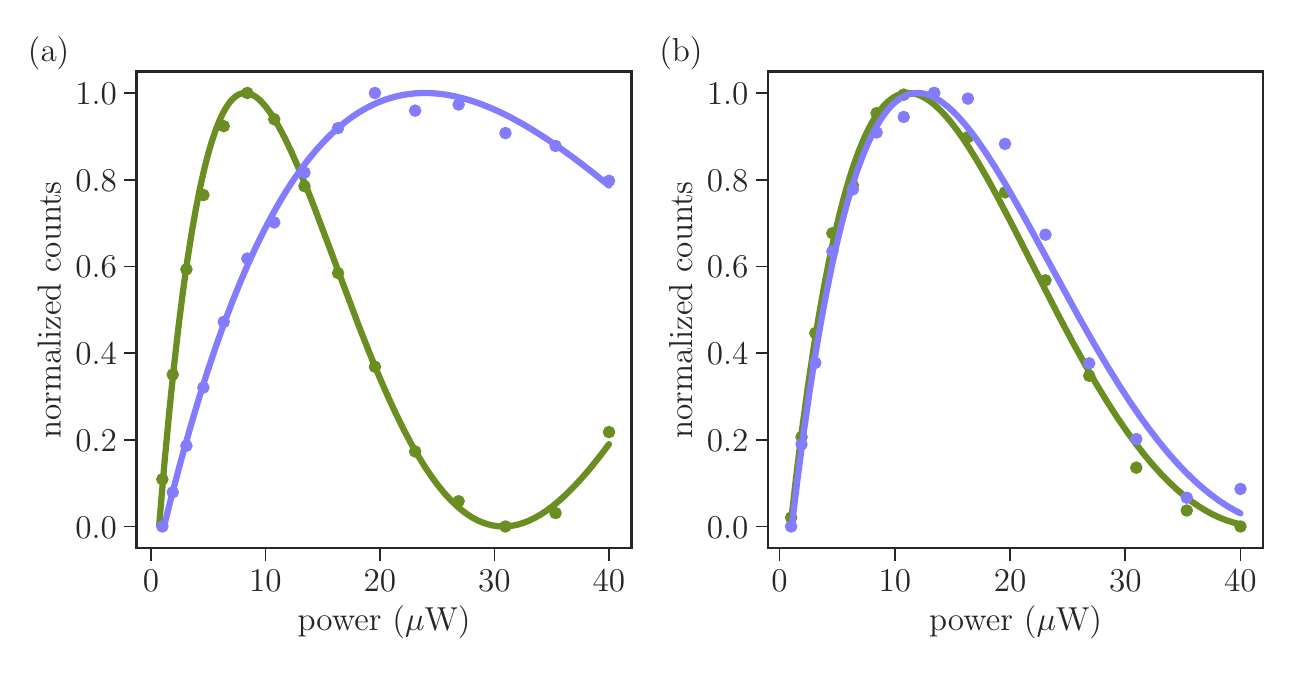}
	\caption{
    Rabi measurements with intensity as a function of excitation power. The driving strengths for QD$_1$ and QD$_2$ are in (a) unequal, and in (b) equal, where the latter is as well shown in \fref{fig:3}(b). Dots are data points and the solid lines are fits, where the color  for QD$_1$ is olive, and for QD$_2$ is purple. 
    }
	\label{fig:SM_Rabicurve}
\end{figure}

Examples of Rabi measurements are shown in \fref{fig:SM_Rabicurve}. For one waveplate configuration, the driving strengths of the two QDs are unequal, as shown in \fref{fig:SM_Rabicurve}(a). After adjustment of the waveplate configuration, the driving strengths are equalized for both dots, as shown in \fref{fig:SM_Rabicurve}(b). The power for a $\pi$-pulse, $P_\pi$, i.e. the power to fully excite from ground to excited state, is extracted from the Rabi oscillation. This extraction is based on the emission intensity being proportional to the population in the excited state, $p_{e,i}$, which for resonant excitation depends on the excitation power, $P$, as
\begin{equation}
p_{e,i} = \sin^2\left( \eta_{exc} \sqrt{P} \right),
\end{equation}
with $\eta_{exc}$ a scaling coefficient that depends on the set-up excitation efficiency. The Rabi frequency, $|\Omega_i|$, is proportional to $\sqrt{P}$, thus the change in $\pi$-pulse power as a function of polarization is a direct measure for the relative driving strength. 

\begin{figure}[h!]
 	\includegraphics[width=.67\columnwidth]{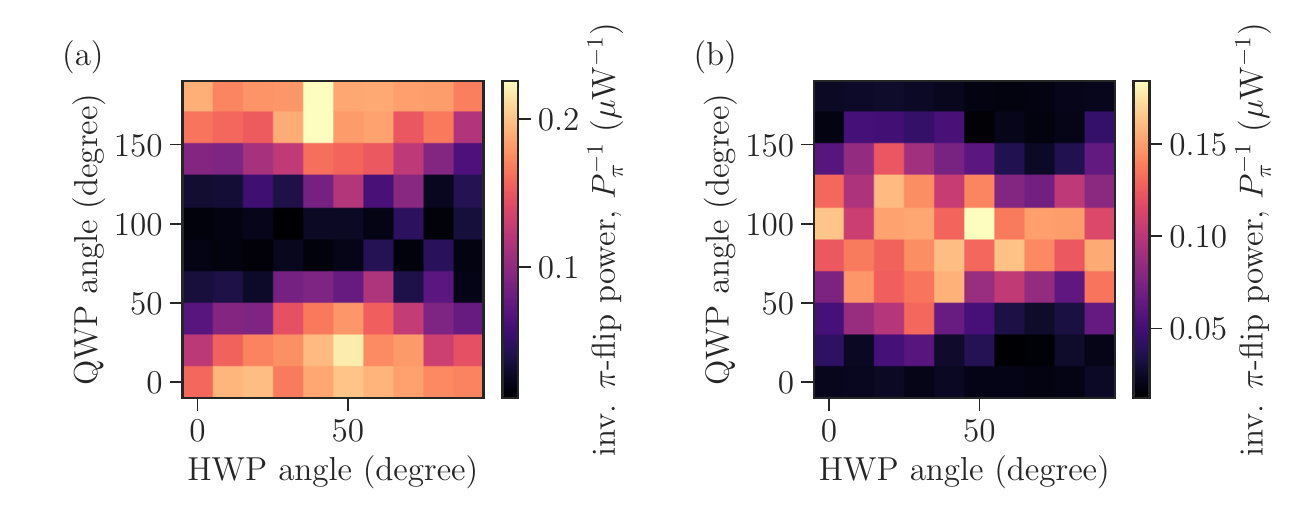}
	\caption{
   (a), (b) The inverse of the $\pi$-pulse power for a Rabi oscillation measurement as a function of quarter-wave plate (QWP) and half-wave plate (HWP) rotation angles for (a) QD$_1$ and (b) QD$_2$. 
   }
	\label{fig:SM_Rabimap}
\end{figure}

Figure~\ref{fig:SM_Rabimap}(a) and (b) show maps of the inverse $\pi$-pulse power as a function of waveplate rotation angles for QD$_1$ and QD$_2$, respectively. These maps were acquired by performing a Rabi measurement for each waveplate setting. Note that the maps in \fref{fig:SM_Rabimap}, which are directly obtained from Rabi oscillation measurements, display the same pattern as the waveplate maps in ~\fref{fig:SM_waveplatemaps}, which were acquired at a fixed power. This agreement confirms that the latter measurement is sufficient to capture the dependence of the Rabi frequency on polarization. 

\end{document}